*Article*

# The Effect of Marketing Investment on Firm Value and Systematic Risk

Musaab Mousa [1], Saeed Nosratabadi [1], Judit Sagi [2,*] and Amir Mosavi [3,4,5,6,*]

[1] Doctoral School of Economic and Reginal Sciences, Szent Istvan University, 2100 Godollo, Hungary; Musaab.Mousa@phd.uni-szie.hu (M.M.), saeed.nosratabadi@phd.uni-szie.hu (S.N.)
[2] Department of Finance, Budapest Business School, 1149 Budapest, Hungary
[3] John von Neumann Faculty of Informatics, Obuda University, 1034 Budapest, Hungary
[4] School of the Built Environment, Oxford Brookes University, Oxford OX3 0BP, UK
[5] Department of Informatics, Selye Janos University, 94501 Komarom, Slovakia
[6] School of Economics and Business, Norwegian University of Life Sciences, 1430 Ås, Norway
* Correspondence: sagi.judit@uni-bge.hu (J.S.); amir.mosavi@mailbox.tu-dresden.de (A.M.)

**Abstract:** Analyzing the financial benefit of marketing is still a critical topic for both practitioners and researchers. Companies consider marketing costs as a type of investment and expect this investment to be returned to the company in the form of profit. On the other hand, companies adopt different innovative strategies to increase their value. Therefore, this study aims to test the impact of marketing investment on firm value and systematic risk. To do so, data related to four Arabic emerging markets during the period 2010–2019 are considered, and firm share price and beta share are considered to measure firm value and systematic risk, respectively. Since a firm's ownership concentration is a determinant factor in firm value and systematic risk, this variable is considered a moderated variable in the relationship between marketing investment and firm value and systematic risk. The findings of the study, using panel data regression, indicate that increasing investment in marketing has a positive effect on the firm value valuation model. It is also found that the ownership concentration variable has a reinforcing role in the relationship between marketing investment and firm value. It is also disclosed that it moderates the systematic risk aligned with the monitoring impact of controlling shareholders. This study provides a logical combination of governance–marketing dimensions to interpret performance indicators in the capital market.

**Keywords:** marketing; investment; emerging markets; firm value; systematic risk; share value; ownership concentration; social science; open innovation; complexity

## 1. Introduction

In terms of performance appraisal and impact assessment strategies, marketing has undergone fundamental changes beyond product-market measures—i.e., market share and sales growth—and addresses capital market measures, such as firm value and share stock return elements [1,2]. Accordingly, based on integration and overlap among economic fields, competition between companies in the product market stretches to the capital market, in the sense that the higher a company's position in the market, the better its performance in the capital market [3]. Likewise, the product-market demand leads to a difference in the prospective returns in the capital market with regard to the competition; all strategic and operating actions of the company interact directly or indirectly with its market value [4]. Additionally, competition is mainly connected to marketing, as the most common competition tool in business, by creating competitive advantages to persuade potential customers to choose the company's products or services without other alternatives available in the market [5]; increasingly, marketing has become more inclined to innovation to overcome challenges and protect stakeholders' interests. Therefore, marketing activities have become the major driver of a company's performance in terms of traditional performance characteristics and improving the returns of shareholders [6]. In the same manner, marketing efforts reflect a long-term investment for a company that may directly bring financial benefits (such as returns and profitability) and indirectly bring marketing benefits (such as customer satisfaction). In turn, such benefits implicitly help to explain market value [7]. As a result of the growing importance of marketing in organizations, many scholars have endeavored to explain the financial impact of marketing strategies through the relationship between marketing efforts and company performance in the capital market [8,9].

The previous interdisciplinary literature ultimately falls under the marketing–finance interface, as a new scientific approach deals with the joint impact of the financial and non-financial elements embodied in the firm's value. In other words, marketers have begun to adopt a vision that considers the financial aspects of marketing strategies in order to contribute to achieving the original firm goal of maximizing the owners' wealth [10]. However, the current radical changes in marketing synchronize with a significant development in the capital market concept, which becomes the essential criteria of the firm performance in the framework of maximizing shareholder value—in other words, transferring value to investors [11]. Hence, the capital market metrics used by researchers to measure the effect of marketing listed firm performance, chiefly share price, as a primary expression of market value and systematic risk, which lies at the core of portfolio theory through the linkage between the performance of the company's stock and the performance of the overall market portfolio in harmony with the market-based asset creation framework, where the investment in marketing leads to generating some intangible assets, such as brand equity and customer equity, which in turn play a significant role in firm value enhancement and relevant risk lowering, depending on cash flow features [12]. Marketing practices accelerate cash flows, which increases the value and supports the stability of revenues—that is, it reduces fluctuations in cash flow and thus reduces risks [13,14]. On the other hand, the various marketing efforts send a clear signal to the capital market, which has a fundamental influence on the investor's response and decisions towards the company's shares, especially the share price and its liquidity level, which reaffirms the long-term nature of marketing investment [15]. Furthermore, the success of marketing in achieving satisfactory financial outputs in relation to firm value depends, to a large extent, on the degree of financial constraints, which explains the difference between countries regarding the marketing–firm value relationship [16].

It is found that most of the relevant empirical studies were performed on developed market frameworks, and most of them also focused on assessing the impact of marketing variables on financial variables using common evaluation models. In this context, by using a sample of the highly trading companies in some Arab emerging markets for the period between 2010 and 2019, the current research aims to analyze the impact of marketing on the capital market—particularly on two metrics. The first is the firm value by proposing a firm valuation model, which involves a marketing investment variable as a complementary element of accounting for published numbers, while the second is analyzing the relationship of systematic risk and marketing investment controlled by size, age, and financial leverage. In addition to testing the moderating role of ownership concentration, which adds a governance dimension to the proposed models, meaning that current research is trying to answer the question of to what extent ownership can increase the validity and predictive power of the evaluation model, this provides new evidence for the literature related to the marketing–finance interface in the framework of emerging markets. Additionally, research findings show that marketing applications could play a significant role in leveraging value and rationalizing investment decisions in emerging markets, on the one hand, and in risk reduction, on the other hand. This, in turn, can be a key element to increase the efficiency of these markets and motivate investors, which leads to a greater contribution to economic development.

The rest of this paper includes the relative literature about the relationship between marketing and performance measured in capital marketing. In the next section, the methodology, the sample and data procedures, and the formulation of the proposed model are provided. Ultimately, the statistical results are displayed with a discussion.

## 2. Research Background

*2.1. Marketing, Firm Value, and Systematic Risk*

Ref. [17] analyzed the trend of corporate cost for 50 years from 1945 to 1995; their results revealed that all elements belonging to manufacturing costs dropped from 30% to 50% as a percentage of total corporate costs; the administrative costs contribution dropped from 30% to 20%; while the trend of marketing costs was reversed, rising from 20% to 50% of total costs over the five decades. Further, the marketing budget average equals 11.2% of the global revenue, ranging between 22% in the retail sector and 2.6% in the health and pharma sector [18].

The shift in marketing expenditure as a long-term investment is an obvious phenomenon in modern business. For example, published financial statements of Apple corporation show $933 million as the marketing expenses against $87.1 billion for the brand value items [19]. The research literature deals with marketing firm value through two paths of marketing variables; the first one focuses on marketing assets' impact as an ultimate outcome of marketing investment, while the second one deals with the impacts of some marketing actions/strategies as the initial inputs of marketing investment [20]. By analyzing a considerable set of empirical studies, ref. [21] concluded that both marketing assets and marketing actions have a clear elasticity through used capital market valuation models; it has been revealed that the elasticity of marketing assets is higher than that from advertising from the marketing actions perspective.

Regarding the impact of marketing assets, brand equity has attracted great interest from researchers, and early attempts to explain the role of a brand were concentrated on its link with the future firm [22]. Relying on the Capital Asset Pricing Model (CAPM) valuation model, ref. [23] shows that a high brand value portfolio benefits from a higher level of return and lower level of risk compared to other companies listed and the market return average in the Turkish market, which was confirmed in Latin American markets where companies included in a valuable brand finance list have a lower risk level and higher return level when compared with their counterparts not on the list [24]. In the Arabic emerging market, [25] shows that brands enhance their share return and have an informative contain to motivate market response.

On the other hand, in the developed markets framework companies with a high brand capital investment and high brand investment per employee gain higher returns [26]. In this regard, the research team concluded that brand value correlates positively to return parameters and negatively to both systemic and idiosyncratic risk embedded on CAPM factors [27]. Furthermore, a high brand value could lower the negative impact of market crises such as the global financial crisis of 2008 [28].

Moreover, within marketing assets' collection customer equity has received a high level of priority in marketing, since the customer is the core of business strategies. Similarly, customer equity as an intangible market asset provided a reasonable proxy for firm value and was characterized as an appropriate approach regardless of the firm lifecycle period, especially during the growth peak or times of negative profit, where the traditional financial models could not be applied smoothly [29]. As well as customer satisfaction, customer loyalty became an efficient measure of companies' strategic success as well as a measure of the financial outputs of marketing [30]. In the same manner, customer measurements such as the Customer Satisfaction Index correlate positively with a firm value from one hand and reduce the cost of capital on the other hand [31]. Customer satisfaction information presents a reliable signal to motivate the investor's response to the company; for example, when Dell's customer satisfaction score went down in August 2005 by 6.3%, the share price dropped by 12.5% [32].

The second part of marketing investment involves marketing actions, which have been interpreted by scholars in the framework of capital market performance. Initially, advertising action is the perceptible part of marketing. It is clear and visible to the audience, and most of the advertising spending information of listed companies is available in popular databases [10]. Thus, a large body of research has addressed the impact of advertising on firm valuation criteria and related risk in the capital market, where an increase in advertising spending leads to less systematic risk and improved financial health [2,33]. Similarly, advertising intensity leads to a low degree of implied cost of capital [34]; this is because of the increase in investors' familiarity level with the company, which in turn leads to a higher level of liquidity and return [35]. Additionally, advertising communication could be an important resource to support investment decisions by providing a clear signal to the company, allowing it to price its products properly and at the same time informing investors about the right value of shares [36]; thus, investors choose stocks with higher advertising, therefore making it possible that the behavior of the investor could be modified by advertising communication [15].

In addition to advertising, new product introduction is considered the most influential marketing action on firm value. Introducing a new product is a major outcome of adopting an innovation approach through monitoring and transferring market feedback into actionable inputs to develop the current product or introduce a new one in light of perceived customer needs, ensuring the stable revenue of the firm or reducing the likelihood of risk [37]. This enhances the long-term value of the firm as a result of the investor's reaction to new available information, which intensifies over time [38], while irregularity in the product introducing process has a negative impact on the firm's value [39]. Initially, a new innovative product explains and motivates the firm's value growth compared with imitative products, leading to a lower level of value growth as measured by the Tobin Q TQ ratio [40]. This extends to a new product announcement, which leads to significant abnormal return, since the announcing of a new product would boost the attractiveness of a firm's traded shares [41].

It is worth mentioning that Beta, as a matrix of systematic risk, despite the fact that it is considered a basic portion of the Capital Asset Pricing Model (CAPM), is also an agreed-upon tool to build an efficient investment portfolio [33]. Besides this, it has been used as a common proxy for the cost of capital in a lot of previous empirical research [2]. Based on what is mentioned above, marketing variables influence capital market metrics, so it is expected that the relationship can be applied in Arab emerging markets. Thus, the first two hypotheses are as follows:

**Hypotheses 1 (H1).** *Marketing investment has a positive significant impact on firm value.*

**Hypotheses 2 (H2).** *Marketing investment has a negative significant impact on firm systematic risk.*

*2.2. The Role of Ownership*

Joint stock ownership structure differs from other corporate legal forms by the nature of ownership, especially in terms of the owner rights as well as its link to capital market mechanisms. Inherently, ownership structure is associated with agency theory, where some conflicts are produced, such as owner–manager conflict and controlling-noncontrolling owner conflict [42]. The implications of the ownership disparity between shareholders are formed in two directions; the first is monitoring impact, which involves the ability of large shareholders to control managers' decisions and thus reduce the possibility of managers harming the interests of shareholders or engaging in opportunistic behavior. The second direction is the expropriation impact, which involves the negative aspect of large shareholder–minority shareholder conflict, assuming that controlling shareholders act in their interest regardless of other owners' interests by transforming recurses and cash flow for their private benefit, which is known as the tunneling phenomenon. In other words, ownership structure is a vital pillar of the corporate governance system [43,44]. Prior studies have dealt with the relationship of firm performance in the capital market and many ownership structure aspects, such as managerial ownership, institutional ownership, bank ownership, and family ownership. It must be noted that studies that have dealt with the direct relationship of marketing elements and ownership are rare in previous literature, except for [35], who concluded that advertising expenditure contributes to an increase in the number of shareholders and thus a high ownership dispersal.

In relation to positive monitoring impact, a plethora of research proves this impact empirically. In [45], the authors conclude on the positive effect of concentrated ownership in terms of firm value based on controlling and minority owners' convergence of interest in the Spanish market. Along the same line, [46] showed that funder-controlled companies perform better in the market than non-funder-controlled companies in China, where funder-concentrated ownership motivates investors by being a firewall for the company from their point of view. Additionally, the ownership percentage of the largest shareholder and the largest three shareholders correlate positively with firm value in Romania [47].

On the other hand, other studies have reemphasized the negative expropriation impact. In [48], the authors demonstrated that more increased control that is not coupled with good cash flows led to lower market value during the Asian crisis. The author of [49] tested the relationship between ownership concentration measured by individually controlling shareholders' percentages and institutionally controlling shareholders and firm value; he found that both measures push down the firm value in Switzerland. Likewise, the expropriation impact of ownership concentration in the Korean market deepens the negative R&D–firm value relationship because of controlling shareholders hindering R&D investment decisions [50]. Meanwhile, a third line of research revealed no clear link between ownership concentration and firm performance; the authors of [51] reported that a high level of control by family or state shareholders in Arab Gulf listed companies did not show a significant impact of ownership on the market to book ratio.

Concerning the systematic risk–ownership nexus, in the light of conflict of roles for different segments of shareholders, the existence of several controlling shareholders increases the market firm risk, while a single controlling shareholder contributes significantly to risk reduction in the USA [52]. It was shown that companies controlled by shareholders who own diversified portfolios tend to take more risks compared with others controlled by non-diversified shareholders in Europe. The same effect was proven in the banking industry, where the controlling shareholders push toward risky decisions to increase their wealth. On the contrary, ref. [53] documented that the ownership concentration has no impact on market risk, as measured by unexpected volatility, and performance, as measured by the TQ ratio, in Vietnam.

In essence, the variation in the results of ownership impact on value and risk is due to the characteristics of the country or region being studied regarding the level of regulatory institution development in relation to governance framework in general and particularly the degree of investor protection.

Consequently, the third and fourth hypotheses are as follows:

**Hypotheses 3 (H3).** *Ownership concentration moderates the relationship between marketing investment and firm value.*

**Hypotheses 4 (H4).** *Ownership concentration moderates the relationship between marketing investment and firm systematic risk.*

The conceptual model of the study is formed according to the literature and the hypotheses. This model is presented in Figure 1.

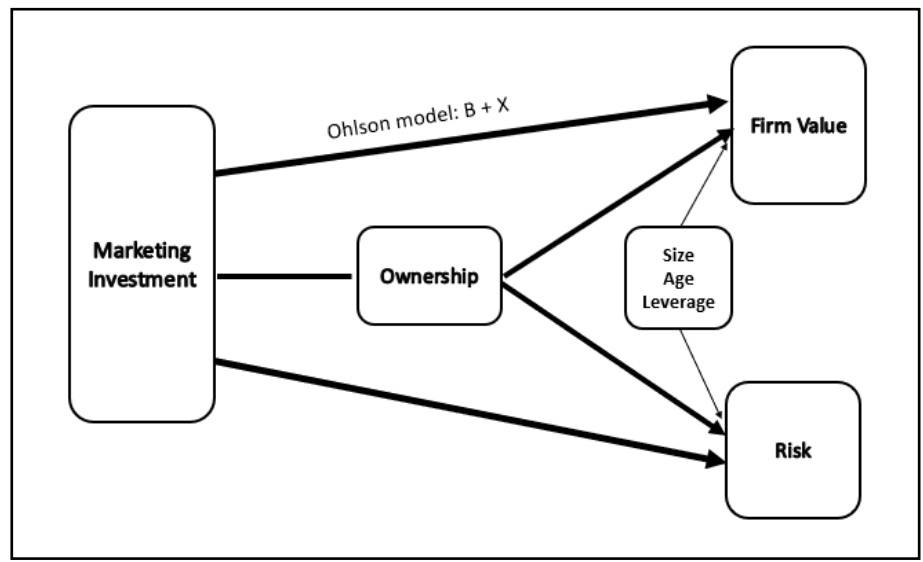

**Figure 1.** Conceptual model of research.

## 3. Methodology

*Data and Model Formation*

Four markets were selected (Qatar, Dubai, Abu Dhabi, Kuwait) based on them having similar economic and social circumstances as well as similar financial market structures. In the next step, the listed companies within the constituents of the Financial Times Emerging Markets Index (FTSEEMI) are identified; they include 44 companies. Specific characteristics were applied to determine the final sample regarding the available financial statements from 2010 to 2019; the companies had a clear product with alternatives in the market and positive book values during the study period. After dropping financial-sector companies, 20 companies (200 observations) were accepted in the final sample, as shown in Table 1, which is based on an adaptation from FTSE Russell (2020). The sample companies present the most-traded listed companies in the markets under study by weight at 1.85% of the FTSE Emerging Markets Index (FTSEEMI).

**Table 1.** Sample of research.

| Variable | Qatar | Dubai | Abu Dhabi | Kuwait | Total |
|---|---|---|---|---|---|
| Market listed companies | 43 | 68 | 70 | 216 | 397 |
| FTSEEMI constituents | 19 | 8 | 7 | 10 | 44 |
| Weight in FTSEEMI | 1.14% | 0.43% | 0.5% | 0.93% | 3% |
| Final sample | 10 | 3 | 3 | 4 | 20 |
| Sample weight in FTSEEMI | 0.9% | 0.21% | 0.31% | 0.43% | 1.85% |
| Market cap USD B | 14.4 | 7.4 | 8.5 | 8.1 | 38.4 |

Furthermore, we used secondary data of fundamental financial figures and data related to sharing the price of sample companies extracted from the Thomson Reuters Refinitiv DataStream as well as from the official websites of markets and companies in the case of missing data.

The current research adopts Ohlson (1995) as one of the most critical residual earnings-based valuation models which was published in 1995; some refinements were applied later [54–57]. Substantially, the model has gained considerable attention among related research due to its logical assumptions and mathematical structure depending on accounting figures. Additionally, one of the most essential advantages of the model is that the firm value is independent of the accounting choices effect.

According to Ohlson (1995), normal earnings are equal to the book value at the previous year $t-1$ multiplied by the cost of capital; for that, abnormal earnings are the output of subtracting normal earnings from actual earnings, as in Equation (1):

$$X_t^a = X_t - rB_{t-1}, \tag{1}$$

where $X_t^a$ = abnormal earnings for period $t$; $X_t$ = earnings per share for period $t$; $r$ = risk-free return; $B_{t-1}$ = book value for period $t-1$.

The model assumes the time series behavior of abnormal earnings through a linear information dynamic, which is considered the most important contribution of the model, as it created a link between current information and intrinsic value according to Equation (2):

$$P_t = B_t + a_1 X_t^a + \beta_1 V_t, \tag{2}$$

where $P_t$ = market value of share for period t; $B_t$ = book value for period t; $X_t^a$ = abnormal earnings per share for period t, which presented in (1); $V_t$ = information other than accounting information.

According to Ohlson (1995), the valuation model expressed in (2) concludes that the abnormal earnings are produced by the company's monopoly position in the product market and that the returns tend towards the cost of capital in the long run due to the competition level. On the other hand, $V_t$ demonstrates that other information determines the price more than accounting information; in other words, other elements could play a significant role in investor decisions. This assumption is harmonious with the marketing firm value research stream in connection with additional information provided by marketing variables to accounting numbers to forecast stock prices [22,31]. Accordingly, current research uses the marketing investment as a proxy for other information in the model which is measured by marketing expenses, calculated as the selling and general administrative expenses (SG&A) minus R&D expenses [58–60]. Due to the role of marketing as a long-term investment, marketing expenses are divided by total sales:

$$Marin_t = [(SG\&A - R\&D)]/Sales. \tag{3}$$

Otherwise, marketing investment variables obtain a comprehensive proxy that takes into account all marketing applications in both marketing assets and marketing actions. Therefore, the main model is presented in Equation (4):

$$P_t = B_t + a_1 X_t^a + a_2 Marin_t, \tag{4}$$

where $Marin_t$ = marketing investment for the period $t$; $X_t^a$ = abnormal earnings for the period $t$.

The Ohlson 1995 model provides a logical framework of market value–residual earning linkage on the one hand and takes the other valuable resources into account on the other hand, particularly the goodwill role in value creation [61], which is in line with the concept of intangible marketing assets as a supplement to the accounting information of tangible assets, which could be an adequate measure to narrow the obvious variation between market value and disclosed accounting information. Marketing efforts can add predictive power to the valuation model in parallel with abnormal earnings, particularly explaining the gap between the market and book value through creating intangible marketing assets which provide a convenient explication of observations related to market value.

On the other hand, to show the individual differences among sample companies, some control variables have been added to the model—namely, company age, as measured by the number of years since establishment, since older companies will have more accumulated intangible assets. The second control variable is company size, as measured by the natural logarithm of total assets at the end of the period. Finally, financial leverage has been added to the model, measured by the total equity to total assets ratio to control the effect of the financial structure of the sample; thus, the direct impact of the model is presented in Equation (5):

$$P_t = B_t + a_1 X_t^a + a_2 Marin_t + a_3 Age + a_4 Size + a_5 Lev + \varepsilon_t. \quad (5)$$

According to the ownership concentration as a moderating variable, following the relative studies concentration calculated by the controlling the shareholders' total ownership percentage on 31st December, the controlling ownership threshold is calculated based on 5% of voting rights [52]. The moderating impact demonstrated in Equation 6 requires generating a new variable for the interaction between the interpreted variable Marin and the moderating variable OW [62].

$$P_t = B_t + a_1 X_t^a + a_2 Marin_t + a_3 Age + a_4 Size + a_5 Lev + a_6 OW + a_7 OW * Marin + \varepsilon_{ti}, \quad (6)$$

where $B_t$ = share book value; $X_t^a$ = abnormal earnings for the period $t$; $Marin_t$ = marketing investment for the period t; Age = firm age; Size = natural logarithm of total assets; Lev = financial leverage; OW = ownership concentration.

On the other hand, this research aims to explain the impact of marketing investment on the related risk of capital market through the systematic risk factor, which is calculated along the lines of related literature [33] based on regression estimation between the equal-weighted monthly return of share and market index using a moving five-year window (60 months or at least 48) Therefore, this research proposes the model in Equation (7) for the risk–marketing investment relationship:

$$Bet_t = c_1 Marin_t + c_2 Age + c_3 Size + c_4 Lev + \varepsilon_{ti} \quad (7)$$

In the same manner, the moderating impact of ownership concentration is demonstrated in Equation (8) with the proposed control variables:

$$Bet_t = c_1 Marin_t + c_2 Age + c_3 Size + c_4 Lev + c_5 OW + c_6 OW * Marin + \varepsilon_{ti}, \quad (8)$$

where $Bet_t$ = systematic risk factor; $Marin_t$ = marketing investment for the period t; $Age$ = firm age; $Size$ = natural logarithm of total assets; $Lev$ = financial leverage; $OW$ = ownership concentration. The definition of the variables is presented in Table 2.

In this session, the statistical results are displayed to test the research hypotheses, starting from some descriptive statistics, then using panel data regression steps to test the relationship between marketing investment, firm value, and systematic risk in the markets under study.

**Table 2.** Definition of research variables.

| Variable | Description | Recourse |
|---|---|---|
| Share price | Annual closing price. | Thomson Reuters DataStream |
| Book value **B** | Book value in 31st December, equity/outstanding share number. | Thomson Reuters DataStream |
| Abnormal earnings X | Earnings per share less the normal earning $rB_{t-1}$ $X_t^a = X_t - rB_{t-1}$, where $r$ = risk-free return measured by the yield of government bonds for ten years. | Own calculation based on Thomson Reuters DataStream |
| Marketing investment **Marin** | Marketing expenses/sales. | Own calculation based on Thomson Reuters DataStream |
| Ownership concentration **OW** | Total ownership percentage of the controlling shareholders (5% of voting right). | Thomson Reuters DataStream |
| Size | Ln (total assets). | Thomson Reuters DataStream |
| Age | Number of years from establishment. | Thomson Reuters DataStream |
| Financial leverage **Lev** | Total equity/total assets. | Thomson Reuters DataStream |
| Systematic risk **Bet** | Systematic risk factor calculated by using a moving five-year window (60 months or at least 48) through regression estimation between the monthly return of the share and the market index $\beta et_i$ = slop ($R_i$, $R_m$). | Own calculation based on Thomson Reuters DataStream |

*4.1. Descriptive Statistics*

Table 3 includes the mean, standard deviation, and maximum and minimum values of sample companies. The minimum price within the sample companies was $0.32, where the share has been traded at a level close to the par value ($0.27), while the maximum price was $6.04 with a mean at $1.70; also, the book value ranges between $0.07 and $2.78, with an average of $1.28, which indicates a good level of retained earnings and hence a greater margin of safety, supporting the stability of the company's financial position. Regarding abnormal earnings X, the main is close to 11%, while the range of portfolio between −40 and 42% could be an acceptable level of performance measured by operational profit. With respect to marketing investment, the average for marketing expenditure was close to 25% of sales, within a range between 0.001 and 43%. The age of the portfolio companies ranges between 13 and 62 years, with an average of 19.68 years, which reflects the relatively short history of the sample companies as an extension of the economic and financial structure of countries under study; this is also relatively recent compared to other regions of the world. With respect to the systematic risk measured by beta, the average of the sample portfolio was 0.89, meaning that the return of the companies under study depends on the market return by 89%, with a maximum value at 240%. Furthermore, the ownership concentration average is 44% in a range from 22% and 90%, which reflects a high level of concentration. Finally, the market to book ratio ranges between 0.47 and 2.45 times with an average at 2.24, which exceeds the global average at 1.7 and that of emerging markets at 1.3 [63]; this refers to the value created by intangible asset investment, particularly marketing assets.

**Table 3.** Descriptive statics of the sample.

| Variable | N | Minimum | Maximum | Mean | Std. Deviation |
|---|---|---|---|---|---|
| P | 200 | 0.32 | 6.04 | 1.7080 | 1.2252 |
| B | 200 | 0.07 | 2.78 | 1.2874 | 1.6355 |
| X | 200 | −0.400 | 0.4201 | 0.1094 | 0.1606 |
| Marin | 200 | 0.001 | 0.4324 | 0.2491 | 0.1565 |
| Age | 200 | 3.00 | 62.00 | 19.6834 | 17.730 |
| Total Assets USD M | 200 | 108 | 259,532 | 20,958 | 30,292 |
| Lev | 200 | 0.0400 | 0.9100 | 0.5288 | 0.2062 |
| Bet | 200 | −0.38 | 2.41 | 0.8931 | 0.3373 |
| OW | 200 | 0.22 | 0.90 | 0.4400 | 0.2213 |
| P/B | 200 | 0.47 | 2,45 | 2.24 | 0.2949 |

*4.2. Correlation Test*

Table 4 shows the correlation coefficients of variables, where the relationship direction corresponds with the study hypothesis. Furthermore, the correlation outputs refer to a significant relationship between the first dependent variable P and the independent variable value of Sig is less than 5% ($p < 0.05$), noting the full correlation of share price with abnormal earnings, which supports the Ohlson model's assumption of a linear time series behavior. Similarly, the systematic risk Bet correlates inversely with most of the independent variables as well as with the market value presented by P.

**Table 4.** Variable correlation matrix.

| Probability | P | X2 | B | MAR | BETA | LEVR | OW | SIZ | AGE |
|---|---|---|---|---|---|---|---|---|---|
| P | 1 | | | | | | | | |
|   | ----- | | | | | | | | |
| X | 0.6359 | 1 | | | | | | | |
|   | 0.0000 | ----- | | | | | | | |
| Marin | 0.0359 | 0.0107 | 0.0258 | 1 | | | | | |
|   | 0.0145 | 0.8804 | 0.7172 | ----- | | | | | |
| Bet | −0.4134 | −0.3434 | −0.3582 | −0.2496 | 1 | | | | |
|   | 0.0000 | 0.0000 | 0.0000 | 0.0004 | ----- | | | | |
| Lev | −0.0215 | 0.0281 | 0.0750 | 0.0314 | -0.0570 | 1 | | | |
|   | 0.7634 | 0.6941 | 0.2926 | 0.6599 | 0.0239 | ----- | | | |
| OW | 0.2238 | 0.1118 | 0.1892 | 0.0543 | −0.1890 | −0.2140 | 1 | | |
|   | 0.0015 | 0.1160 | 0.0074 | 0.4461 | 0.0075 | 0.2400 | ----- | | |
| Size | −0.1187 | −0.0474 | 0.0792 | 0.1054 | 0.2542 | −0.3989 | 0.2468 | 1 | |
|   | 0.0949 | 0.5065 | 0.2660 | 0.1383 | 0.0003 | 0.0900 | 0.4004 | ----- | |
| Age | 0.1638 | 0.0617 | 0.3315 | −0.0724 | −0.0196 | 0.2001 | 0.2157 | 0.3769 | 1 |
|   | 0.0208 | 0.3870 | 0.0230 | 0.3095 | 0.0837 | 0.0046 | 0.0022 | 0.0701 | ----- |

*4.3. Model Estimation Procedure*

Due to the structure of the study data, a panel data analysis was conducted by running the model estimation to define the fit model, which reflects the relationship between dependent and independent variables depending on relative statistics. Firstly, heteroskedasticity checking was conducted to test the consistency of the residual error variance in variables data based on the Likelihood Ratio (LR). Hence, we used a white cross-section within the covariance method to control the adverse heteroskedasticity impact on the model estimation. The next step was the stationarity checking of the models' variables by the Unit Root Test (URT); the stationarity is the initial term of model estimation to avoid the misleading regression of time series. URT was conducted using Augmented Dickey-Fuller (ADF) (1979), as in Table 5.

**Table 5.** Stationarity URT test of variables.

| Variable | Level | 1 Difference | 1 Difference | Order |
|---|---|---|---|---|
| P | −7.1288 | - | - | I(0) |
| X | −6.4318 | - | - | I(0) |
| Marin | −8.0179 | - | - | I(0) |
| Age | −3.8236 | - | - | I(0) |
| Size | −1.8097 | - | - | I(0) |
| Lev | −7.3267 | - | - | I(0) |
| Bet | −1.7010 | - | - | I(0) |
| OW | −8.0179 | - | - | I(0) |

From Table 5, all the variables show stationarity in their level form I(0), where the prob value is less than 5%, meaning that the null hypothesis of unit root is rejected and all the variables are integrated on the level I(0)

*4.4. The Estimation of Marketing Investment Impact on Firm Value*

The panel data model estimation of firm value as a dependent variable including the variable coefficients and the related statistics of model significance are presented in Table 6. After conducting the Hausman test (1978) to choose the appropriate method of model estimation, the resulting prob value of Chi-Squared is less than 5%, so the null hypothesis is rejected and the fixed-effects method is fit to the research data.

**Table 6.** Marketing investment and firm value: estimation results.

| Dependent: P | Direct Model | | Moderating Model | |
| --- | --- | --- | --- | --- |
| Variable | Coefficient | Prob. | Coefficient | Prob. |
| C | 0.8056 | 0.0000 | 0.8083 | 0.0030 |
| X | 3.0205 | 0.0774 | 2.9186 | 0.0421 |
| Marin | 0.0219 | 0.3843 | 0.1823 | 0.0452 |
| AGE | 0.0145 | 0.0201 | 0.0118 | 0.0330 |
| Size | −0.3376 | 0.0012 | −0.3504 | 0.0014 |
| Lev | −1.7540 | 0.0001 | −1.3985 | 0.0003 |
| OW | | | 0.1880 | 0.0005 |
| OW*Marin | | | 0.1143 | 0.0328 |
| R-squared | 0.4273 | | 0.6047 | |

Table 6 includes the fixed effects regression results of the Ohlson model with a white cross-section to avoid heteroskedasticity impact on share price as a dependent variable. We include marketing investment as a proxy for other information and control variables, as well as the moderating impact of ownership concentration OW*Marin over the period 2010–2019.

The table shows that the dependent variable is affected significantly by abnormal earnings and marketing investment is based on the prob value, which is less than 5%. The explanatory power is based on an R-squared value of 0.42, which means that the independent variables explain 42% of the share price variance; on the other hand, the model is fit for estimation based on the F-statistic, which is less than 5%. Furthermore, the ownership concentration moderates the firm value–marketing investment relationship, where the prob value of the coefficient is less than 5% for both the OW variable and the interaction variable OW*Marin, knowing that the explanatory power has increased significantly, moderating the model at 60%. This confirms the monitoring impact of controlling shareholders on performance, supporting our first and second hypotheses.

*4.5. The Estimation of Marketing Investment Impact on Systematic Risk*

The result of the Hausman test (1978) indicates that the fixed-effects method is more suitable for risk model estimation (Chi-Squared Prop ≤ 5%); the estimation outputs of both direct and ownership moderating impact are shown in Table 7.

**Table 7.** Marketing investment and systemic risk: estimation results.

| Dependent: Bet | Direct Model | | Moderating Model | |
| --- | --- | --- | --- | --- |
| Variable | Coefficient | Prob. | Coefficient | Prob. |
| C | −0.8144 | 0.0213 | −0.7271 | 0.0001 |
| Marin | −0.1962 | 0.0000 | −0.6494 | 0.0012 |
| AGE | −0.0101 | 0.0012 | −0.0098 | 0.0100 |
| SIZ | 0.2043 | 0.0000 | 0.1881 | 0.0055 |
| LEVR | 0.5588 | 0.0175 | 0.6956 | 0.0047 |
| OW | | | −0.2573 | 0.0310 |
| OW*Marin | | | −1.7233 | 0.0202 |
| R-squared | 0.1933 | | 0.3421 | |

Table 7 includes the fixed-effects regression results of the marketing investment impact on systematic risk Bet with a white cross-section to avoid heteroskedasticity impact and control variables, as well as the moderate impact of ownership concentration OW*Marin over the period 2010–2019.

The table shows that marketing investment affects reversely the systemic risk Bet, meaning that the marketing investment expenditure is lower the related risk of the listed company, which extends to all variables in the model except size and leverage, where the larger the company, the larger the level of associated risk that is a result of the high integration of sized firms into economic factors. Similarly, high financial leverage raises the level of exposure to risk due to debt service pressures. On the other hand, ownership concentration leads to a lower level of systemic risk, whereas dispersed ownership may increase the potential for interest conflicts and thus exposure to external risks, especially when the block shareholders have long-term investment goals in maintaining relative stability at the price level and mitigating fluctuations. Moreover, the ownership variable moderates the impact of marketing investment on systemic risk based on the Prop value of the coefficient, which is less than 5% for both the OW variable and the interaction variable OW*Marin. Otherwise, the ownership moderating role enhances the explanatory power of the model from 19% to 34%. This supports our third and fourth hypotheses.

*4.6. Robustness Test*

To ensure the robustness of our baseline findings, the two alternatives of marketing investment measures have been used in panel data regression to test the reliability of the statistical outputs in both the firm value and systematic models. Firstly, the marketing expenses to total assets ratio has been employed as a proxy of marketing investment; second, the natural logarithm of marketing expenses has been used alternatively, as shown in Table 8, where the outputs of alternatives measures correspond to our basic results in relation to the marketing impact on firm value in the first part of the table, as well as the impact of marketing investment on systemic risks in the second part of the table, where the outputs of alternative measures confirm the direct impact of marketing investment on firm value as well as the moderating impact of ownership on the valuation model. This is extended regarding the reversed impact of marketing investment on systemic risk.

**Table 8.** Marketing investment alternatives: estimation results.

| Dependent: P | First Alternative | | Second Alternative | |
|---|---|---|---|---|
| Variable | Coefficient | Prob. | Coefficient | Prob. |
| C | 1.095751 | 0.0000 | 1.177912 | 0.0101 |
| X | 1.553398 | 0.0005 | 1.177912 | 0.0000 |
| Marin (1) | 1.224370 | 0.0000 | 0.105217 | 0.0351 |
| AGE | −0.010499 | 0.0090 | −0.003231 | 0.2122 |
| Size | −0.055548 | 0.0007 | −0.046980 | 0.0495 |
| Lev | −0.186262 | 0.0764 | −0.539324 | 0.0202 |
| OW | 1.784478 | 0.0000 | 2.159767 | 0.0116 |
| OW*Marin | −29.72153 | 0.0000 | −0.345049 | 0.0038 |
| R-squared | 0.361569 | | | |
| Dependent: Bet | First Alternative | | Second Alternative | |
| Variable | Coefficient | Prob. | Coefficient | Prob. |
| C | 0.0042 | 0.0001 | 0.00146 | 0.0014 |
| Marin (1) | −0.035610 | 0.1734 | 0.000356 | 0.0173 |
| AGE | −0.004720 | 0.0494 | −0.004720 | 0.0494 |
| SIZ | 0.139299 | 0.0000 | 0.139299 | 0.0000 |
| LEVR | −0.002469 | 0.9848 | −0.002469 | 0.9848 |
| OW | −0.200487 | 0.3141 | −0.200487 | 0.3141 |
| OW*Marin | −0.001194 | 0.0071 | −0.001194 | 0.0071 |
| R-squared | 0.290261 | | 0.271341 | |

Table 8 includes the fixed effects regression results of marketing investment alternatives impact on the firm value P and systematic risk Bet with white cross-section to avoid heteroskedasticity impact and control variables, as well as the moderating impact of ownership concentration OW*Marin over the period 2010–2019.

## 5. Discussion: Marketing Investment, Firm Value, and Open Innovation

*5.1. Marketing Investment and Firm Value*

The insight of the marketing role in capital market indicators has turned into a new trend among researchers and practitioners. The research tries to demonstrate the impact of marketing on firm value in the emerging markets context, therefore the statistical results provide interesting insight into the role of marketing in shareholder value generation. The results are consistent with relevant literature regarding the positive relationship between marketing expenditure and financial performance, which in turn reemphasizes the growing importance of marketing strategies as a driver of performance enhancement [64]. Furthermore, firm value could be an inclusive metric to measure performance, since it involves many factors influencing both the internal and external environment of the business, and wealth maximizing is the goal on which all stakeholders agree [65].

The results provide new evidence about the reciprocal influence between product market and capital market, where the marketing activities are an efficient channel to transfer the impact of product market elements such as customer reaction and competition level, which in turn translates to value in the capital market.

From other side, the results confirm the reliability of the Ohlson model in firm value valuation depending on residual earning, meaning that the firm value of the sample portfolio is a function of the share book value and abnormal earning, providing new evidence of the Ohlson model's significance, where the firm value is inherently determined by the investor's expectation about the firm's future through comparing the accounting earnings with their investment costs [66]. At the same time, adding marketing variables as a proxy for other information factors enhances the valuation model's power. That is, the marketing variable boosts the informative content of accounting figures—in other words, marketing information plays a complementary role in stimulating investor response.

The results are consistent with the role of market-based assets in value creation as an outcome of marketing investment, as reported by [12]. The positive effect on cash flows is presented by residual earnings in the proposed model as a logic channel to improve firm value, which could by a practical approach to interpreting the growing contribution of intangible assets as a pivot portion of firm value. This means that the markets under study evaluate the intangible marketing assets that lead to more excellent market value [67], and using marketing investment could be an efficient path by which to evaluate the intangible assets as a step forward to frame the numerical recognition of this type of asset.

In relation to systemic risk, our findings indicate that increases in marketing investment lead to a lower level of systemic risk, in line with previous empirical research [2,33], and marketing investment-systematic risk reconciliation provide a deeper vision into the role of the non-financial factor in market reaction to firm shares, and thus the contribution to shareholders value creation that can be elucidated as marketing applications promotes firm value and lessens its linkage to market trend simultaneously through the inverse relationship in the proposed model, in the sense that marketing investment helps relatively in highlighting the individual investment features of a firm's share in isolation from the market movement.

The ownership concentration plays a significant role in reinforcing the impact of marketing investment on firm value that is aligned to a monitoring assumption where large owners use their voting power to curb the opportunistic behavior by motivating managers to make operational decisions aiming at increasing firm value and in turn increasing their fortunes [43]. In addition, the power of large shareholders can be extended to marketing decisions such as the adoption of strategies and agreed on budgets, knowing that the sample companies are characterized as concentrated in terms of ownership. Generally, the positive influence of concentrated shareholders on financial performance and particularly on capital market measures has been proven by a relative stream of studies [45,53,68]. On the other hand, our findings do not agree with those of prior studies of a positive ownership–risk relationship, which is classified under the expropriation role of controlling shareholders [52]. Therefore, based on our findings, this role of ownership can be discussed in terms of how the controlling shareholders contribute to mitigating the potential risk or systematic risk as a core element of a listed company, which could be a result of the nature of block shareholders in the market under study, as a strategic investor aims to maintenance their investment value at an acceptable return by avoiding high volatility in their portfolio price, especially in the case of state shareholders. The marketing–ownership combination would add an operational dimension represented in the outputs of marketing strategies and an administrative dimension as represented by ownership concentration to provide a deeper and more comprehensive interpretation of the firm performance and its value drivers.

The findings indicate that marketing investment reduces the risk associated with the company's capital market activities, especially in emerging markets that have higher risk due to economic and political uncertainty. Therefore, an efficient marketing investment company can ensure a more stable price of their financial assets in the capital market [24].

*5.2. Marketing Investment and Open Innovation*

The marketing–finance interface is an operational reflection of open innovation and market complexity through its distinctive determinants. In this context, the new marketing approach is a focal point for promoting open innovation of the business; marketing could be a bridge to exchange the company's internal knowledge with external parties as well as help build a relationship framework that allows the optimal investment of the company's competitive advantages [69]. Further, marketing monitors the available information in the market to use it for developing new products or services and contributes to rationalizing innovation strategies and ensuring the company's competitive position. Particularly from the perspective of new product acceptance and adoption by the customer, where the customer prefers a product developed by an open innovation process which accelerates cash flow/revenue and reduces fluctuation/risk, this justifies the allocation of a portion of R&D budget to capture and develop ideas from outside resources [70].

Likewise, open innovation adoption can be an effective path for building marketing assets, especially by applying a modern arrangement such as a virtual brand community where customer groups play an outstanding role in brand image enhancement through interactive value co-creation. In other words, open innovation can be a driver in the value accumulation of intangible marketing assets. The effect mechanism of open innovation in a firm valuation workflow is illustrated by the integration of open innovation into the marketing value chain, which in turn activates the inherent organizational capacity of the business and contributes to increasing the effectiveness of the marketing investment to maximize the market value in order to meet the expectations of various stakeholders.

**6. Conclusions**

This study investigated the direct impact of marketing investment on firm value and systematic risk. The present study was conducted on the most traded companies in four Arabic emerging markets, and ownership concentration was considered as a moderating variable. The findings point to the positive effect of marketing investment on firm value through the promoting role of the ownership variable, while marketing investment has a negative effect on systematic risk. These findings contribute to the research literature on the framework of the marketing–finance interface. This study provides evidence about informative content marketing elements and developing the valuation model of the Ohlson model. Indeed, the study's proposed model enriches the debate about the reliability of marketing actions and their role as a long-term investment in shareholders' value. On the other hand, the results related to ownership concentration highlight the importance of ownership structure mechanisms in enhancing governance, particularly in emerging markets. This is why the governance increases the degree of marketing investment efficiency in market value creation. Capital markets, especially emerging markets, face high levels of risk due to economic and political uncertainty. The findings of the current study reveal that marketing investments are able to reduce such risks in emerging markets. Thus, it is recommended that companies should think about effective investment in marketing because it will result in a more stable price for their assets in the capital market.

For future research, by using the interdisciplinary methodology, more variables could be studied in the light of the marketing–firm valuation relationship, and considering other variables as a proxy for firm value or performance could enhance the analysis results, in addition to analyzing the potential applications of open innovation in leveraging the marketing role in performance. Furthermore, other ownership structure elements could be analyzed to show their individual impact. Finally, the Arab markets are not deep enough in terms of the number of listed companies and the eligible companies for listing in the Emerging Markets Index, which has reduced the number of sample items. Therefore, it is strongly recommended to conduct more empirical studies covering a larger number of listed companies.

**Author Contributions:** Conceptualization, M.M. and J.S.; methodology, M.M., S.N., A.M.; software, M.M.; validation, J.S., A.M.; investigation, M.M.; writing—original draft preparation, M.M. and S.N.; writing—review and editing, M.M., S.N., and A.M.; supervision, J.S., A.M. All authors have read and agreed to the published version of the manuscript.

**Funding:** Support of Alexander von Humboldt foundation is acknowledged.

**Acknowledgments:** Support of Alexander von Humboldt foundation is acknowledged.

**Conflicts of Interest:** The authors declare no conflict of interest.

## References

1. Joshi, A.; Hanssens, D.M. The Direct and Indirect Effects of Advertising Spending on Firm Value. *J. Mark.* **2010**, *74*, 20–33, doi:10.1509/jmkg.74.1.20.
2. Singh, M.; Faircloth, S.; Nejadmalayeri, A. Capital Market Impact of Product Marketing Strategy: Evidence from the Relationship Between Advertising Expenses and cost of capital. *J. Acad. Mark.* **2005**, *33*, 432–444.
3. Jory, S.; Ngo, T. Firm power in product market and stock returns. *Q. Rev. Econ. Financ.* **2017**, *65*, 182–193, doi:10.1016/j.qref.2016.09.008.
4. Aguerrevere, F.L. Real Options, Product Market Competition, and Asset Returns. *J. Financ.* **2009**, *64*, 957–983, doi:10.1111/j.1540-6261.2009.01454.x.
5. Burnett, J. Core Concepts of Marketing, The Global Text Project. 2008. Available online: https://www.saylor.org/site/wp-content/uploads/2012/11/Core-Concepts-of-Marketing.pdf (accessed on 2 December 2020)
6. Weber, J.A. Managing the marketing budget in a cost constrained environment. *Ind. Mark. Manag.* **2002**, *31*, 705–717, doi:10.1016/s0019-8501(01)00191-2.
7. Tudose, M.P.; Alexa, L. The effect of marketing expenses on car sales–an empirical analysis. *MATEC Web Conf.* **2017**, *126*, 1–4.
8. Cavusgil, S.T.; Zou, S. Marketing Strategy-Performance Relationship: An Investigation of the Empirical Link in Export Market Ventures. *J. Mark.* **1994**, *58*, 1–21.
9. Agic, E.; Cinjarevic, M.; Kurtovic, E.; Cicic, M. Strategic marketing patterns and performance implications. *Eur. J. Mark.* **2016**, *50*, 2216–2248, doi:10.1108/ejm-08-2015-0589.
10. Zinkhan, G.M.; Verbrugge, J.A. the marketing/finance interface: Two divergent and complementary views of the firm. *J. Bus. Res.* **2000**, *50*, 139–142.
11. Sacui, V.; Dumitru, F. Market-based assets. Building value through marketing investments. *Procedia Soc. Behav. Sci.* **2014**, *124*, 157–164.
12. Srivastava, R.K.; Tasadduq, A.; Liam, F. Market-Based Assets and Shareholder Value: A Framework for Analysis. *J. Mark.* **1998**, *62*, 2–18.
13. Rao, V.R.; Agarwal, M.K.; Dahlhoff, D. How is Manifest Branding Strategy Related to the Intangible Value of a Corporation? *J. Mark.* **2004**, *68*, 126–141, doi:10.1509/jmkg.68.4.126.42735.
14. Anderson, E.W.; Fornell, C.; Mazvancheryl, S.K. Customer Satisfaction and Shareholder Value. *J. Mark.* **2004**, *68*, 172–186, doi:10.1509/jmkg.68.4.172.42723.
15. Frieder, L.; Subrahmanyam, A. Brand Perceptions and the Market for Common Stock. *J. Financ. Quant. Anal.* **2005**, *40*, 57–85.
16. Palomino-Tamayo, W.; Juan, T.; Cerviño, J. The Firm Value and Marketing Intensity Decision in Conditions of Financial Constraint: A Comparative Study of the United States and Latin America. *J. Int. Mark.* **2020**, *28*, 21–39.
17. Sheth, J.N.; Sisodia, R.S. Feeling the heat: Marketing is under fire to account for what it spends. *Mark. Manag.* **1995**, *4*, 8–23.
18. Statista. Global Marketing Spending 2010–2019. 2020. Available online: www.statista.com (accessed on 2 December 2020).

19. Badenhausen, K. Apple Tops List of the World's Most Powerful Brands. *Forbes* **2012**, *10*. Available online: https://www.forbes.com/sites/kurtbadenhausen/2012/10/02/apple-tops-list-of-the-worlds-most-powerful-brands/#22a2b63ecc9a (accessed on 11 November 2020).
20. Srinivasan, S.; Hanssens, D. Marketing and firm value: Metrics, methods, mindings, and future directions. *J. Mark. Res.* **2009**, *XLVI*, 293–312.
21. Edeling, A.; Fischer, M. Marketing's Impact on Firm Value: Generalizations from a Meta-Analysis. *J. Mark. Res.* **2016**, *53*, 515–534, doi:10.1509/jmr.14.0046.
22. Aaker, D.A.; Jacobson, R. The Value Relevance of Brand Attitude in High-Technology Markets. *J. Mark. Res.* **2001**, *38*, 485–493, doi:10.1509/jmkr.38.4.485.18905.
23. Bank, S.; Yzar, E.; Sivri, U. The portfolios with strong brand value: More returns? Lower risk? *Borsa Istanb. Rev.* **2019**, *20*, 64–79, doi:10.1016/j.bir.2019.09.001.
24. Oliveira, M.O.; Stefanan, A.; Lobler, M. Brand equity, risk and return in Latin America. *J. Prod. Brand Manag.* **2018**, *27*, 557–572, doi:10.1108/jpbm-02-2017-1418.
25. Mousa, M.; Sági, J.; Zéman, Z. Brand and Firm Value: Evidence from Arab Emerging Markets. *Economies* **2021**, *9*, 5, doi:10.3390/economies9010005.
26. Belo, F.; Lin, X.; Vitorino, M.A. Brand capital and firm value. *Rev. Econ. Dyn.* **2014**, *17*, 150–169.
27. Fehle, F.; Fournier, S.M.; Madden, T.J.; Shrider, D.G. Brand value and asset pricing. *Q. J. Financ. Account.* **2008**, *47*, 3–26.
28. Johansson, J.K.; Dimofte, C.V.; Mazvancheryl, S.K. The performance of global brands in the 2008 financial crisis: A test of two brand value measures. *Int. J. Res. Mark.* **2012**, *29*, 235–245.
29. Gupta, S.; Lehmann, D.; Stuart, J. Valuing customers. *J. Mark. Res.* **2004**, *41*, 7–18.
30. Gupta, S.; Valarie, Z. Customer Metrics and Their Impact on Financial Performance. *Mark. Sci.* **2006**, *25*, 687–717.
31. Sullivan, D.; Mccallig, J. Does customer satisfaction influence the relationship between earnings and firm value? *Mark. Lett.* **2009**, *20*, 337–351.
32. Fornell, C.; Mithas, S.; Morgeson, F.V.; Krishnan, M.S. Customer Satisfaction and Stock Prices: High Returns, Low Risk. *J. Mark.* **2006**, *70*, 3–14.
33. McAlister, L.; Srinivasan, R.; Kim, M. Advertising, Research and Development, and Systematic Risk of the Firm. *J. Mark.* **2007**, *71*, 35–48.
34. Huang, Y.; Wei, S.X. Advertising intensity, investor recognition, and implied cost of capital. *Rev. Quant. Financ. Account.* **2012**, *38*, 275–298, doi:10.1007/s11156-011-0228-1.
35. Grullon, G.; Kanatas, G.; Weston, J.P. Advertising, breadth of ownership and liquidity. *Rev. Financ. Stud.* **2004**, *17*, 439–461.
36. Chemmanur, T.J.; Yan, A. Advertising, Attention, and Stock Returns. *Q. J. Financ.* **2019**, *9*, 1–51.
37. Prifti, R.; Alimehmeti, G. Market orientation, innovation, and firm performance—An analysis of Albanian firms. *J. Innov. Entrep.* **2017**, *6*, 1–19, doi:10.1186/s13731-017-0069-9.
38. Pauwels, K.; Silva-Risso, J.; Srinivasan, S.; Hanssens, D.M. New products, sales promotions, and firm value: The case of the automobile industry. *J. Mark.* 2004, 68,142–156.
39. Sharma, A.; Saboo, A.R.; Kumar, V. Investigating the Influence of Characteristics of the New Product Introduction Process on Firm Value: The Case of the Pharmaceutical Industry. *J. Mark.* **2018**, *82*, 66–85, doi:10.1509/jm.17.0276.
40. Koski, H.; Kretschmer, T. New product development and firm value in mobile handset production. *Inf. Econ. Policy* **2010**, *22*, 42–50, doi:10.1016/j.infoecopol.2009.11.003.
41. Mann, B.; Babbar, S. New product announcement: Spokesperson a manifestation of increasing firm value. *Int. J. Emerg. Mark.* **2018**, *13*, 1635–1655, doi:10.1108/ijoem-03-2016-0056.
42. Jensen, M.C.; Meckling, W.H. Theory of the firm: Managerial behavior, agency costs and ownership structure. *J. Financ. Econ.* **1976**, *3*, 305–360.
43. Agrawal, A.; Mandelker, G.N. Large Shareholders and the Monitoring of Managers: The Case of Antitakeover Charter Amendments. *J. Financ. Quant. Anal.* **1990**, *25*, 143–161, doi:10.2307/2330821.
44. Claessens, S.; Djankov, S.; Lang, L. The separation of ownership and control in East Asian corporations. *J. Financ. Econ.* **2020**, *58*, 81–112.
45. De Miguel, A.; Pindado, J.; De La Torre, C. Ownership structure and firm value: New evidence from Spain. *Strat. Manag. J.* **2004**, *25*, 1199–1207, doi:10.1002/smj.430.
46. Wang, Q.; Wong, T.J.; Xia, L. State ownership, the institutional environment, and auditor choice: Evidence from China. *J. Account. Econ.* **2008**, *46*, 112–134, doi:10.1016/j.jacceco.2008.04.001.
47. Vintilă, G.; Gherghina, Ș.C. The Impact of Ownership Concentration on Firm Value. Empirical Study of the Bucharest Stock Exchange Listed Companies. *Procedia Econ. Financ.* **2014**, *15*, 271–279, doi:10.1016/s2212-5671(14)00500-0.
48. Nigel, D.; Sarmistha, P. *How Does Ownership Structure Affect Capital Structure and Firm Value? Recent Evidence from East Asia, CEDI Discussion Paper Series 07-04*; Centre for Economic Development and Institutions (CEDI), Brunel University: Uxbridge, UK, 2007.
49. Jentsch, V. Board Composition, Ownership Structure and Firm Value: Empirical Evidence from Switzerland. *Eur. Bus. Organ. Law Rev.* **2019**, *20*, 203–254, doi:10.1007/s40804-018-00128-6.
50. Kang, M.; Kim, S.; Cho, M. The Effect of R&D and the Control–Ownership Wedge on Firm Value: Evidence from Korean Chaebol Firms. *Sustainability* **2019**, *11*, 2986, doi:10.3390/su11102986.

51. Martinez-Garcia, I.; Basco, R.; Gomez-Anson, S.; Boubakri, N. Ownership concentration in the Gulf Cooperation Council. *Int. J. Emerg. Mark.* **2020**, Ahead of Print, doi:10.1108/ijoem-03-2020-0290.
52. Dhillon, A.; Rossetto, S. Ownership Structure, Voting, and Risk. *Rev. Financ. Stud.* **2015**, *28*, 521–560.
53. Tran, N.; Le, D.C. Ownership Concentration, Corporate Risk-Taking and Performance: Evidence from Vietnamese Listed Firms. *Cogent Econ. Financ.* **2020**, *8*, 1–41.
54. Ohlson, J.A. Earnings, book values, and dividends in equity valuation. *Contemp. Account. Res.* **1995**, *11*, 661–687.
55. Ohlson, J.A. Earnings, book values, and dividends in equity valuation: An empirical perspective. *Contemp. Account. Res.* **2001**, *18*, 107–120.
56. Ohlson, J.A. On Accounting-Based Valuation Formulae. *Rev. Account. Stud.* **2005**, *10*, 323–347, doi:10.1007/s11142-005-1534-4.
57. Ohlson, J.A. Accounting Data and Value: The Basic Results. *Contemp. Account. Res.* **2009**, *26*, 231–259, doi:10.1506/car.26.1.8.
58. Mizik, N.; Jacobson, R. Myopic Marketing Management: Evidence of the Phenomenon and Its Long-Term Performance Consequences in the SEO Context. *Mark. Sci.* **2007**, *26*, 361–379, doi:10.1287/mksc.1060.0261.
59. Luo, X. When marketing strategy first meets wall street: Marketing spending and firms' initial public offerings (IPOs). *J. Mark.* **2008**, *72*, 98–109.
60. Ryoo, J.; Jeon, J.; Lee, C. Do marketing activities enhance firm value? Evidence from M&A transactions. *Eur. Manag. J.* **2016**, *34*, 243–257, doi:10.1016/j.emj.2015.11.004.
61. Richardson, G.; Tinaikar, S. Accounting based valuation models: What have we learned. *Account. Financ.* **2004**, *44*, 223–255, doi:10.1111/j.1467-629x.2004.00109.x.
62. Dawson, J.F. Moderation in Management Research: What, Why, When, and How. *J. Bus. Psychol.* **2014**, *29*, 1–19, doi:10.1007/s10869-013-9308-7.
63. Star Capital. Global Stock Market Valuation Ratios. 2020. Available online: https://www.starcapital.de/en/research/stock-market-valuation/ (accessed on 2 November 2020).
64. Auh, S.; Merlo, O. The power of marketing within the firm: Its contribution to business performance and the effect of power asymmetry. *Ind. Mark. Manag.* **2012**, *41*, 861–873, doi:10.1016/j.indmarman.2011.09.021.
65. Rust, R.T.; Ambler, T.; Carpenter, G.; Kumar, V.; Srivastava, R.K. Measuring Marketing Productivity: Current Knowledge and Future Directions. *J. Mark.* **2004**, *68*, 76–89, doi:10.1509/jmkg.68.4.76.42721.
66. Barniv, R.; Myring, M. An International Analysis of Historical and Forecast Earnings in Accounting-Based Valuation Models. *J. Bus. Financ. Account.* **2006**, *33*, 1087–1109, doi:10.1111/j.1468-5957.2006.00596.x.
67. Bond, S.; Cummins, J. The Stock Market and Investment in the New Economy: Some Tangible Facts and Intangible Fictions. *Brook. Pap. Econ. Act.* **2000**, *2000*, 61–107, doi:10.1353/eca.2000.0003.
68. Gaur, S.S.; Bathula, H.; Singh, D. Ownership concentration, board characteristics and firm performance: A contingency framework. *Manag. Decis.* **2015**, *53*, 911–931.
69. Quesado, P.; Silva, R. Activity-Based Costing (ABC) and Its Implication for Open Innovation. *J. Open Innov. Technol. Mark. Complex.* **2021**, *7*, 41, doi:10.3390/joitmc7010041.
70. Liu, X.; Fang, E. Open Innovation: Is it a Good Strategy in Consumers' Eyes? *Acad. Mark. Stud. J.* **2017**, *21*.